\documentclass[
    twocolumn,
]{aastex7}

\usepackage{microtype}

\providecommand{\sorthelp}[1]{}

\begin{document}

\title{Searching for the Shortest-wavelength Aromatic Infrared Bands: \\ No Evidence for the Predicted 1.05\,$\mu$m~Polycyclic Aromatic Hydrocarbon Feature} 

\author[0000-0002-3455-1826, gname=Dennis, sname=Lee]{Dennis Lee}
\affiliation{Jet Propulsion Laboratory, California Institute of Technology, 4800 Oak Grove Drive, Pasadena, CA 91109, USA}
\altaffiliation{\textcopyright~2026 California Institute of Technology. Government sponsorship acknowledged.}
\email[hide]{dennisl@jpl.nasa.gov}

\author[0000-0001-7449-4638, gname=Brandon, sname=Hensley]{Brandon S. Hensley}
\affiliation{Jet Propulsion Laboratory, California Institute of Technology, 4800 Oak Grove Drive, Pasadena, CA 91109, USA}
\email[hide]{brandon.s.hensley@jpl.nasa.gov}

\author[0000-0001-5929-4187]{Tzu-Ching Chang}
\affiliation{Jet Propulsion Laboratory, California Institute of Technology, 4800 Oak Grove Drive, Pasadena, CA 91109, USA}
\affiliation{California Institute of Technology, 1200 E. California Boulevard, Pasadena, CA 91125, USA}
\email[hide]{tzu-ching.chang@jpl.nasa.gov}

\author[0000-0001-7432-2932, gname=Olivier, sname=Dore]{Olivier Dor\'e}
\affiliation{Jet Propulsion Laboratory, California Institute of Technology, 4800 Oak Grove Drive, Pasadena, CA 91109, USA}
\affiliation{California Institute of Technology, 1200 E. California Boulevard, Pasadena, CA 91125, USA}
\email[hide]{olivier.p.dore@jpl.nasa.gov}
    

\begin{abstract}
Polycyclic aromatic hydrocarbons (PAHs) are responsible for a variety of near- and mid-infrared spectral features in Galactic and extragalactic sources.
A feature at 1.05~$\mu$m arising from electronic transitions in PAH cations is predicted by laboratory experiments but has never been observationally confirmed.
We conduct a dedicated search for this feature in absorption on a highly-extinguished sight line toward BD+40~4223, a blue supergiant in Cyg OB2, using the TripleSpec spectrograph at Palomar Observatory.
We place a $5\sigma$ upper limit on the feature strength of $\Delta\tau_{1.05}/A_{V} < 5.6 \times10^{-3}$, ruling out theoretical estimates with $> 10\sigma$~significance.
We constrain the effective temperature of BD+40~4223 to be $\log_{10}\left(T_{\rm eff}\right)=4.41\pm0.03$ and infer that it is veiled by $6.39\pm0.05$ magnitudes of visual extinction, consistent with but more constraining than previous determinations.
As dust on the sight line toward BD+40~4223 appears typical of the diffuse interstellar medium, this non-detection challenges existing models of PAH material properties and/or charge distribution.
\end{abstract}

\section{Introduction}\label{sec:intro}

Polycyclic aromatic hydrocarbons (PAHs) are ubiquitous molecules found throughout the Milky Way as well as in other galaxies. 
These carbonaceous molecules represent the smallest dust grain constituents of the interstellar medium and play a variety of important roles in its thermodynamics and chemistry~\citep{2008ARA&A..46..289T}.

Observationally, PAH vibrational modes give rise to prominent emission features at near- and mid-infrared wavelengths (e.g., 3.3, 6.2, 7.7, 8.6, 11.2~$\mu$m).
The brightness of these features can be used to trace the abundance of PAHs and---when compared to the dust continuum emission---the PAH mass fraction~\citep{2007ApJ...657..810D, planck2013-p06b, 2021ApJ...917....3D}. 
The features have thus proved vital to understanding the properties of dust in a variety of environments in the interstellar medium~\citep[e.g.,][]{2022A&A...666L...5G, 2024A&A...689A..92S, 2025ApJ...994...61L}.
Furthermore, as these features are bright around young massive stars, PAH emission has been used as a tracer of star formation~\citep[e.g.,][]{2004ApJ...613..986P, 2016ApJ...818...60S}.
Studies with the James Webb Space Telescope (JWST), for example, have made extensive use of the near- and mid-infrared emission from PAHs to identify distant star-forming galaxies \citep{2023ApJ...946L..40L} and to trace star formation in nearby ones \citep{2023ApJ...944L..12C}.

A feature that has been included in models of interstellar dust, but has yet to be observationally detected, is the PAH feature centered at 1.05~$\mu$m with a full width at half maximum (FWHM) of 0.058~$\mu$m.
This feature is predicted based on laboratory measurements of electronic transitions in PAH cations~\citep{2005ApJ...629.1188M}.
Since ionized PAHs are believed to {be} common in the interstellar medium~\citep[see][Figure~9]{2001ApJS..134..263W}, \citet{2007ApJ...657..810D} included the 1.05~$\mu$m feature in their model of PAH emission.
Although the feature has not been detected, subsequent dust models have continued to include it in the ionized PAH absorption cross sections~\citep[e.g.,][]{2021ApJ...917....3D, 2023ApJ...948...55H}.

Detecting the 1.05~$\mu$m~feature in emission is difficult.
It is predicted to be approximately two orders of magnitudes weaker than the 3.3~$\mu$m feature and three orders of magnitudes weaker than the 6.2~$\mu$m feature~\citep{2007ApJ...657..810D, 2023ApJ...948...55H}.
Even more importantly, however, is the strength of the interstellar radiation field (ISRF) in the near-infrared.
At $\sim1~\mu$m, interstellar emission is dominated by starlight. 
For conditions typifying the solar neighborhood (ISRF strength $U = 1$, gas column density $N_{\rm H}=2.7\times 10^{21}$  cm$^{-2}$), the~\citet{2007ApJ...657..810D}~model predicts that the 1.05~$\mu$m~emission feature is more than two orders of magnitude fainter than starlight \citep[see][Figure~12.1]{2011piim.book.....D}.
As a result, it is unlikely that the 1.05~$\mu$m PAH feature can be detected in emission.

The most likely avenue for detection of the 1.05~$\mu$m PAH feature is thus in absorption.
The observed depth of the absorption depends only on the intrinsic strength of the feature and the abundance of its carrier \citep{2005ApJ...629.1188M, 2023ApJ...948...55H}.
A sight line with a large quantity of dust would enable detection of, or an upper limit on, this PAH feature despite its relatively weak nature.

In this work, we target the massive star BD+40~4223~($m_J=6.035$ mag)~in the Cyg~OB2 association to constrain the properties of the 1.05~$\mu$m PAH feature~\citep[][]{2006AJ....131.1163S}.
BD+40 4223 (alternatively 2MASS J20323904+4100078) is a spectral type B0\,Ia luminous blue supergiant with an effective temperature $\log_{10}(T_{\rm eff})=4.49$ obscured by $6.53$ magnitudes of visual extinction~\citep{2012AA...543A.101C, 2015MNRAS.449..741W}.
Using near-infrared spectroscopy, we employ a model of the stellar emission, the continuum extinction, and the 1.05~$\mu$m feature to derive a strict non-detection of the feature with strong upper limits inconsistent with the theoretical prediction: $\Delta\tau_{1.05}/A_{V} =(9.5\times10^{-6}) \pm (1.1\times10^{-3})$.
This non-detection challenges our theoretical models as well as demonstrates a need for further laboratory measurements of electronic transitions in PAHs.

This paper is organized as follows:
Section~\ref{sec:observations} details the near-infrared spectrum acquired with the TripleSpec spectrograph. 
Section~\ref{sec:model} and~\ref{sec:results} describe our model and analysis results, respectively.
In Section~\ref{sec:discussion}, we discuss the context and potential implications of the feature's non-detection.
We conclude in Section~\ref{sec:conclusion}.

\section{Observations}\label{sec:observations}

The near-infrared spectrum for BD+40~4223 was acquired with the TripleSpec spectrograph on the Hale 200-inch telescope (P200) at Palomar Observatory on 2025 July 10.
TripleSpec is a cross-dispersed near-infrared spectrograph that covers 0.90~to 2.46\,$\mu$m~with a spectral resolution ($R=\lambda/\Delta\lambda$) of $2700$~\citep{2008SPIE.7014E..0XH}.
BD+40~4223 was observed using the ABBA dithering mode totaling {1418} s of on source time.
In addition, we observed the A0\,V star HD~201320 for use as our telluric standard.

The data were processed and reduced using the IDL-based Spextool software~\citep{2004PASP..116..352V, 2004PASP..116..362C}.
Using Spextool, we performed wavelength calibration, sky removal using the A and B images, order identification and finally extraction of the spectrum.
Following the procedures described in \citet{2003PASP..115..389V}, we then performed telluric absorption correction and flux calibration using observations of HD~201320.

{
For high signal-to-noise spectra (S/N~$> 200$), Spextool has been noted to underestimate the error~\citep{2014AJ....147..160M, 2014AJ....147...20N}.
In this regime, the error is dominated by correlated noise (such as from the telluric correction) rather than random photon noise.
Since our spectra has S/N~$> 300$ throughout most of the range, we account for this by multiplying the Spextool reported uncertainties by a factor of 10. 
This factor is chosen to match the typical point-to-point scatter observed in the spectrum. 
We use this inflated error for our modeling of the extinguished blackbody and 1.05~$\mu$m feature~(Section~\ref{subssec:bb105model}).
}

\section{Model}\label{sec:model}

\subsection{Line Modeling and Subtraction}\label{subsec:linemodel_subtraction}

\begin{figure*}[!ht]
    \centering
    \includegraphics[width=0.99\linewidth]{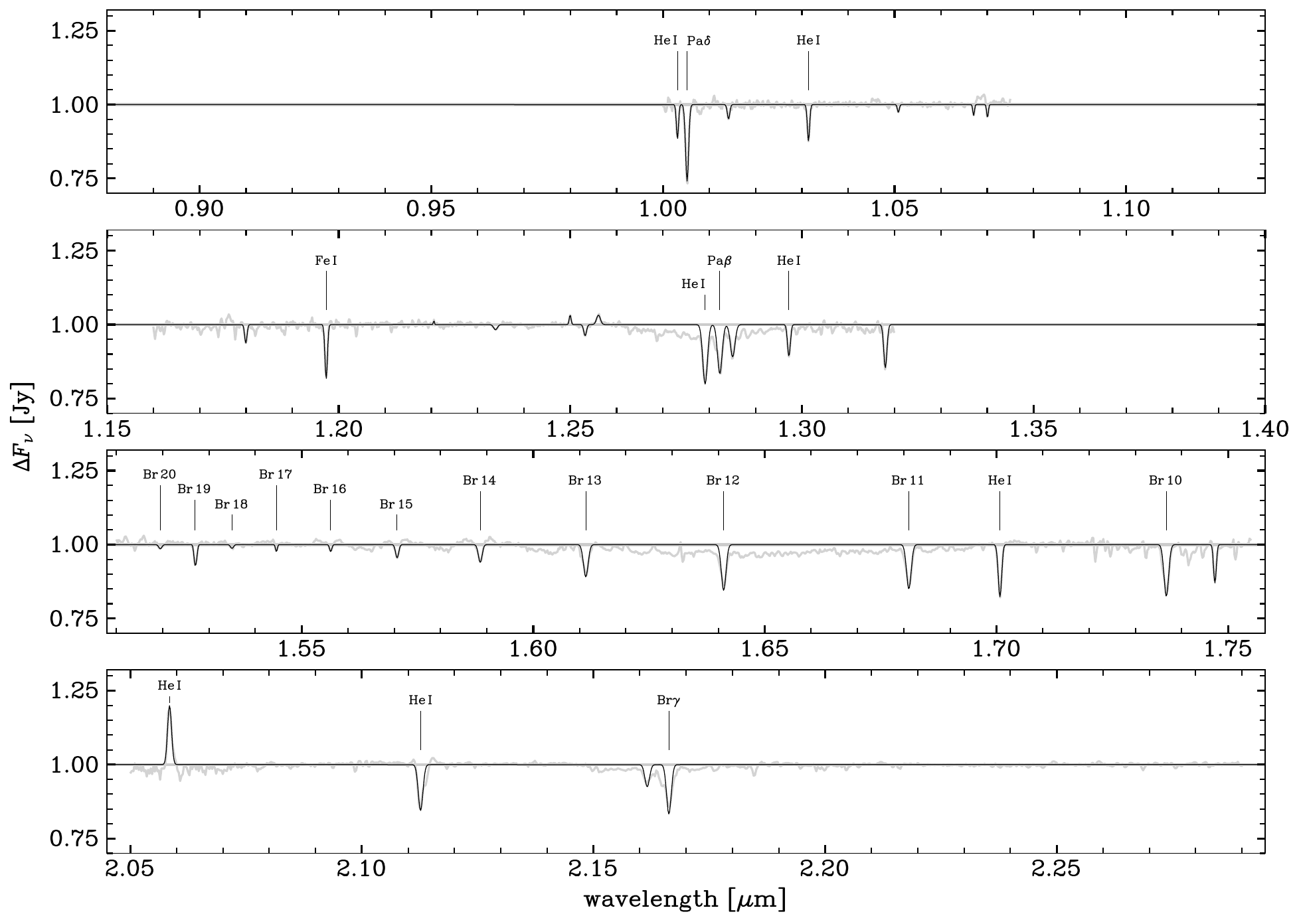}
    \caption{
    Absorption and emission lines in the TripleSpec observations that are fit and subsequently subtracted from the observed spectrum following the procedure in Section~\ref{subsec:linemodel_subtraction}.
    In addition to numerous hydrogen recombination lines---mostly in the Brackett series---these also include various helium lines and an iron line.
    Unidentified lines are unlabelled but are still modeled and subtracted. 
     \label{fig:line-fitting}}
\end{figure*}

While the primary emission is from the stellar atmosphere, numerous hydrogen recombination and helium lines are also seen in the near-infrared spectrum.
As the focus of our analysis is to understand the underlying stellar continuum and its extinction by the intervening dust, we first characterize and subtract these spectral lines from our observations.
Similar to the procedure in \citet{2020ApJ...895...38H}, rather than physically model the complex radiative transfer of the stellar atmosphere of a massive star, we use a simpler approach to subtract the spectral lines.

We perform a spline fit to model the underlying continuum and subtract it from the observed spectrum. 
We then model each line with a Gaussian profile and perform a simultaneous least-squares fit for the amplitude and full width at half maximum (FWHM) of all lines identified by visual inspection.
The {spline-fit continuum subtracted spectrum} and model fit for each line is shown in Figure~\ref{fig:line-fitting}.
We emphasize that the primary goal of this procedure is not a precise characterization and understanding of the observed emission and absorption lines.
Instead, we simply require a sufficient modeling of the lines for subtraction to permit characterization of the continuum. 

\begin{figure*}[!t]
    \centering
    \includegraphics[width=0.99\linewidth]{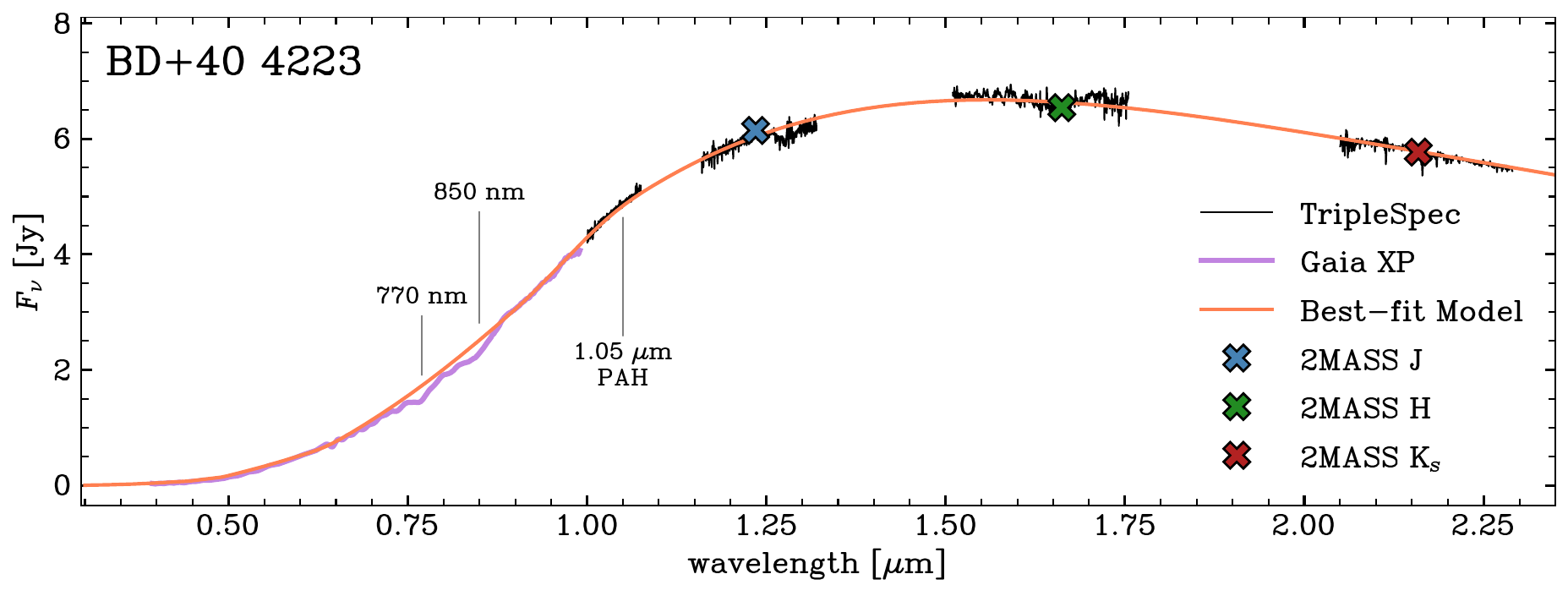}
    \caption{
        TripleSpec observations, shown in black, for BD+40~4223. 
        Regions of the high atmospheric contamination have been masked.
        Visually identified emission and absorption lines have been subtracted following the procedure in Section~\ref{subsec:linemodel_subtraction}.
        The best fit model spectrum---fit only to the TripleSpec observation---is shown as the orange line. 
        The Gaia XP spectrum is shown in purple~\citep{2016A&A...595A...1G, 2023A&A...674A...1G, 2023A&A...674A...2D, 2023A&A...674A...3M}. 
        We also plot measured 2MASS near-infrared fluxes in $J$, $H$, and $K_{\rm S}$ band as the blue, green, and red crosses, respectively~\citep{2006AJ....131.1163S}. 
        The locations of the 770~nm, 850~nm, and 1.05~$\mu$m features are indicated.
        \label{fig:all-data}
    }
\end{figure*}

Figure~\ref{fig:all-data} shows the resulting line-subtracted TripleSpec spectra for BD+40~4223 along with the measured 2MASS near-infrared fluxes in $J$, $H$, and $K_{\rm S}$ band~\citep{2006AJ....131.1163S} and the optical Gaia XP spectrum from 392~nm to 992~nm~\citep{2016A&A...595A...1G, 2023A&A...674A...1G, 2023A&A...674A...2D, 2023A&A...674A...3M}.

\subsection{
    \texorpdfstring{Extinguished Blackbody and 1.05~$\mu$m Feature}{Extinguished Blackbody and 1.05 micron Feature}
}
\label{subssec:bb105model}

For the continuum of our BD+40~4223, we use a simple extinguished blackbody with the inclusion of the expected 1.05\,$\mu$m PAH feature.
We model the continuum observations of our star as:
\begin{equation}\label{eq:model}
    F_\nu^{\rm obs} =  \pi \Omega_* ^2 \cdot B_\nu(T_{\rm eff}) \cdot e^{-\tau(\lambda)}
\end{equation}
where 
$\Omega_* \equiv {R_0}/{D}$ and $R_0$ is the radius of the star and $D$ is the distance to the star.
The extinction is modeled as a combination of a typical extinction curve $\tau_{\lambda, \rm cont}$ and the contribution from the 1.05~$\mu$m feature $\Delta\tau$:
\begin{equation}
    \tau(\lambda) = \tau_{\lambda, \rm cont} + \Delta\tau(\lambda)
\end{equation}
For $\tau_{\lambda, \rm cont}$, we use the \citet{2023ApJ...950...86G} $R_V$-dependent dust extinction model with $R_V=3.1$ \citep[see also][]{2009ApJ...705.1320G, 2019ApJ...886..108F, 2021ApJ...916...33G, 2022ApJ...930...15D}.
To represent a PAH~feature in our model, we use a Drude profile as is commonly used to model dust features \citep[see e.g.,][]{2007ApJ...656..770S}: 
\begin{equation}
    \Delta\tau(\lambda) = C\frac{\gamma^2}{\left(\lambda/\lambda_0 -\lambda_0/\lambda\right)^2 + \gamma^2}~.
\end{equation}
Here, $\lambda_0$ is the central wavelength, $\gamma$ is the fractional FWHM, and $C$ is amplitude at $\lambda_0$.
Following \citet{2007ApJ...657..810D}, for the 1.05\,$\mu$m~feature, we set $\gamma=0.055$ and $\lambda_0=1.050$\,$\mu$m.
As the depth of the feature is expected to scale with extinction, the primary quantity of interest is $\Delta\tau_{1.05}/A_{V}\equiv C/A_V$ where $A_V=1.086\,\tau_V$. 

We use \texttt{emcee}, an affine-invariant Markov chain Monte Carlo (MCMC) ensemble sampler, to find the best-fit parameters~\citep{2013PASP..125..306F}.
We fit for the star's effective temperature ($\log_{10}T_{\rm eff}$), the dust extinction in magnitudes ($A_V$), the normalization factor ($\log_{10}\Omega_{*}$), and the peak depth of the 1.05\,$\mu$m PAH feature per $A_V$ ($\Delta\tau_{1.05}/A_{V}$).

We choose broad uniform priors for all our parameters: 
\begin{eqnarray}
    \log_{10}(T_{\rm eff})~&\sim&~\mathcal{U}(3, 8), \nonumber \\
    \log_{10}(\Omega_*)~&\sim&~\mathcal{U}(-11, -7), \nonumber \\
    A_V~&\sim&~\mathcal{U}(1, 20), \nonumber  \\
    \Delta\tau_{1.05}/A_{V}~&\sim&~\mathcal{U}(0, 10) \nonumber
\end{eqnarray}
For $\Delta\tau_{1.05}/A_{V}$, in our convention, we restrict it to positive values only to ensure that the feature is strictly in absorption.
We perform the MCMC run with 8 walkers and 50,000 iterations and discard the initial 5,000 samples as the burn-in.

\section{Results and Analysis}\label{sec:results}

\begin{deluxetable}{cccc}
\caption{BD+40~4223 Best-fit and Literature Values \label{tab:bestfitparams}}
\tablecolumns{6}
\tablehead{
\colhead{Parameter} &
\colhead{Best-fit Value} &
\colhead{Literature Value} &
\colhead{Ref.}
}
\startdata
$\log_{10}(T_{\rm eff})$         & \phm{$-$}4.41 $\pm$ 0.03 & 4.49 $\pm$ 0.04 & [1] \\ 
$A_V$                  & \phm{$-$}6.39 $\pm$ 0.05 & 6.53 $\pm$ 0.5\phn & [1] \\
$\log_{10}(\Omega_*)$  & $-9.30$ $\pm$ 0.02 &  - & - \\
$\Delta\tau_{1.05}/A_{V}$             & $\left(9.5\pm14 \right)\times10^{-6}$ & 0.017 & [2]
\enddata
\tablerefs{[1]~{\citet{2012AA...543A.101C}}; [2]~\citet{2023ApJ...948...55H}}
\end{deluxetable}

\begin{figure*}
    \centering
    \includegraphics[width=0.95\linewidth]{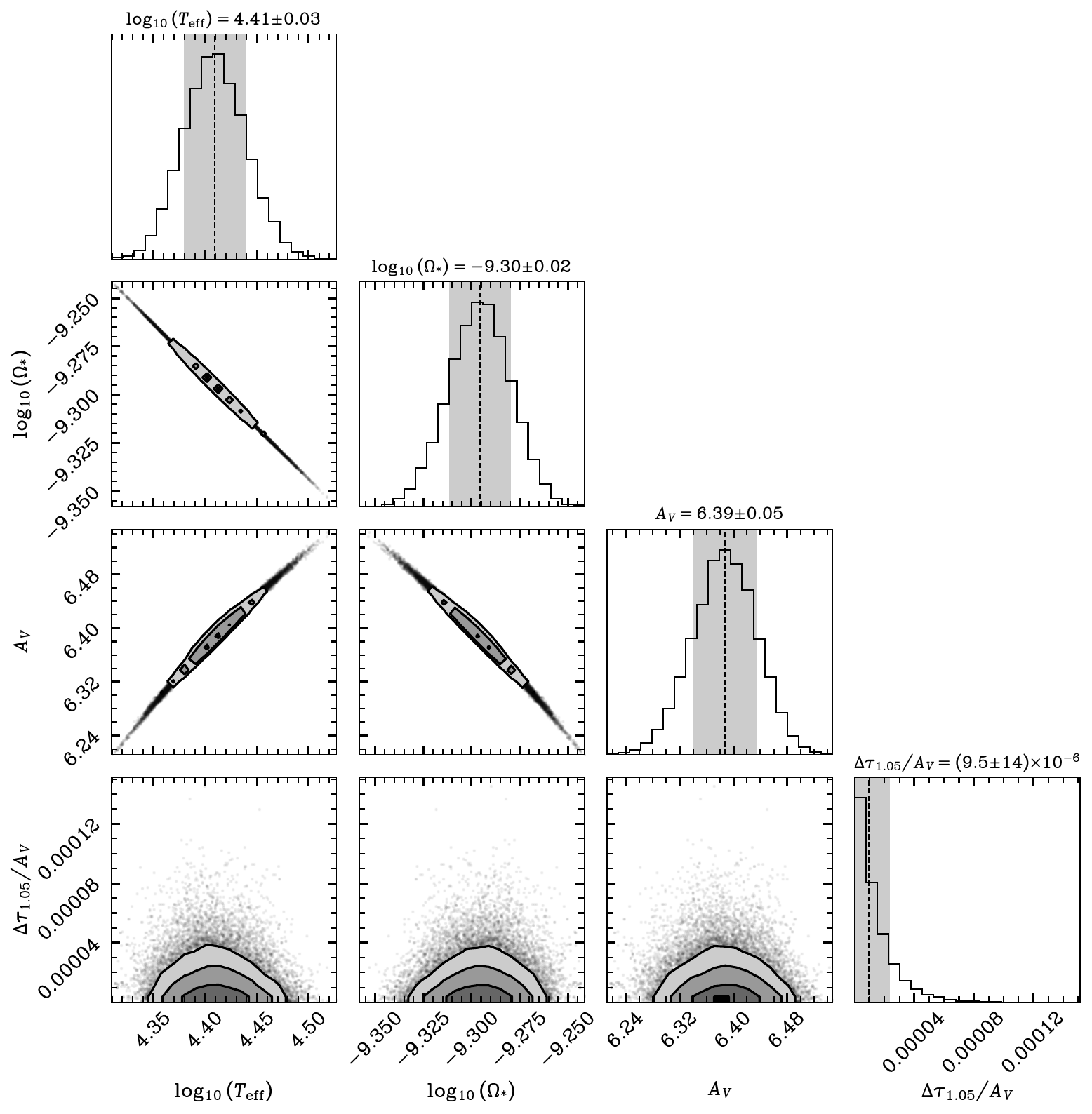}
    \caption{
         The posterior distributions from the MCMC sampling using \texttt{emcee} for BD+40 4223.
         The best fit model (listed in Table~\ref{tab:bestfitparams}) is shown as as the dashed line and the grey band.
        \label{fig:bd404223_corner} 
    }
\end{figure*}

The results of the MCMC sampling are shown in the corner plot in Figure~\ref{fig:bd404223_corner}.
{We verify the convergence of our MCMC by examining the autocorrelation of our sampled chains.
We compute an autocorrelation time of  $\tau_{\rm ACT}=406$. 
As $\tau_{\rm ACT} \times 50$ is less than our number of samples, this indicates that our chains have sufficiently converged~\citep{2013PASP..125..306F}.}

\begin{figure}[!t]
    \centering
    \includegraphics[width=1\linewidth]{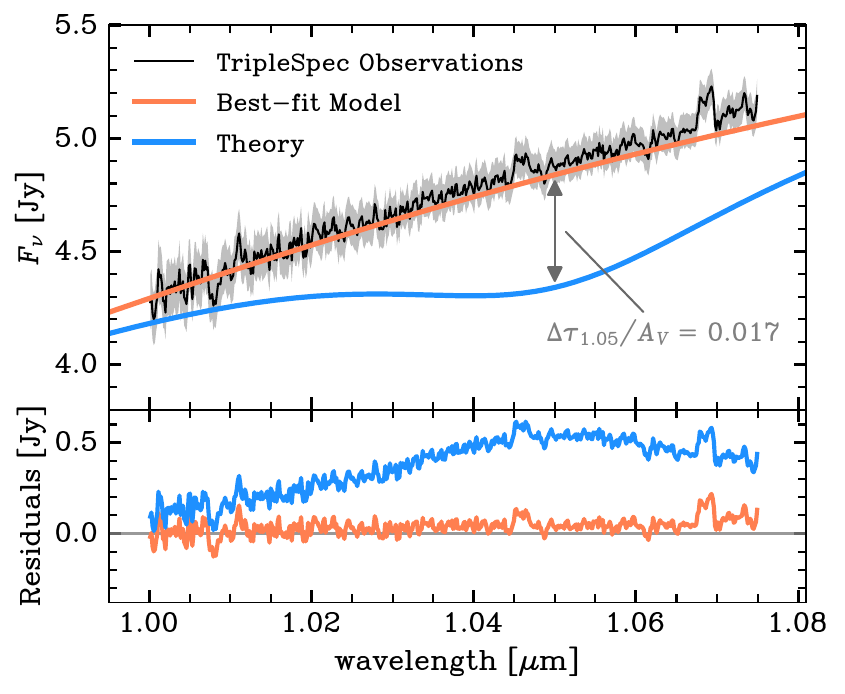}
    \caption{TripleSpec observations, shown in black with $1\sigma$ uncertainties shown in gray. 
    The best-fit model spectrum, fit to the TripleSpec observations, is shown as the orange line. 
    The blue curve indicates a model sharing the same extinction and stellar parameters as our best fit model spectrum, but with the depth of the 1.05 $\mu$m feature is set to the theoretical model from \citet{2023ApJ...948...55H}. The residuals are shown in the lower panel.
    \label{fig:bd40_105}}
\end{figure}

Table~\ref{tab:bestfitparams} lists the median and standard deviation for each marginalized distribution.
Literature stellar parameters for BD+40~4223 and a theoretical prediction for the 1.05\,$\mu$m feature are also listed for comparison.
We use the median of the posterior as our ``best-fit'' model for the rest of the analysis and discussion.
{For this best-fit model, we report a $\chi^2 = 4674$ for our 3408 degrees of freedom. 
This corresponds to a reduced chi-squared of $\chi^2_\nu=1.37$.}
The best-fit model is shown plotted in Figure~\ref{fig:all-data}.
Figure~\ref{fig:bd40_105} also shows the best-fit model in the vicinity of the predicted 1.05~$\mu$m feature.

Our model fit results in a non-detection with a strong upper limit on the strength of the 1.05~$\mu$m feature $\Delta\tau_{1.05}/A_{V}$.
{Based on the posterior distribution, the $5\sigma$ upper limit on the strength of the feature is $\Delta\tau_{1.05}/A_{V} < 7 \times 10^{-5}$.}
In constructing the astrodust+PAH dust model, \citet{2023ApJ...948...55H} predicted the strength of the 1.05\,$\mu$m PAH feature to be $\tau/A_V = 0.017$.
\citet{2005ApJ...629.1188M} also presented an estimate for the absorption strength of the 1.05\,$\mu$m PAH feature $\Delta\tau_{1.05} / E(B-V) = 0.04$ ($\Delta\tau_{1.05}/A_{V} = 0.013$, assuming $R_V = 3.1$).
As this approximation from \citet{2005ApJ...629.1188M} is a simple estimate and since both values are similar, we adopt the value from \citet{2023ApJ...948...55H} as our literature value for comparison.
Figure~\ref{fig:bd40_105} shows our observations in the wavelength range of the putative 1.05~$\mu$m~feature along with our best fit model and the theoretical prediction from \citet{2023ApJ...948...55H}.
{Comparing our measurement of $\Delta\tau_{1.05}/A_{V} =\left(9.5\pm14 \right)\times10^{-6}$ and the theoretical prediction of 0.017, this corresponds to a formal significance of $\approx1200\sigma$. }

Our resulting values for visual extinction $A_V$ and effective temperature $T_{\rm eff}$ are consistent with those reported in the literature from \citet{2012AA...543A.101C}.
\citet{2012AA...543A.101C} derived the effective temperature values by mapping the spectral type on to effective temperature using the calibrations from \citet{2005A&A...436.1049M} and \citet{2000asqu.book.....C} and estimated an uncertainty of 0.04 dex.
The extinction---with a reported uncertainty of 0.5 mag---was estimated by comparing the 2MASS $J-K_{\rm S}$ color with intrinsic color from \citet{2006A&A...457..637M} and \cite{2000asqu.book.....C}.
Given these uncertainties in the prior estimates, our values are statistically consistent.

Using our best-fit value for $\log_{10}\left(\Omega_*\right)$ and the distance toward BD+40~4223 derived from Gaia parallax as $D=1.78\pm0.06$~kpc, we compute a stellar radius of $R_0 = 39.9\pm2.0\,R_\odot$.
The prototypical B0\,Ia supergiant---$\varepsilon$~Orionis---has been estimated to have a radius up to $32.4\,R_\odot$~\citep{1990PASP..102..379W, 2008A&A...481..777S, 2020A&A...643A..88Z}.
Our estimate for the radius of BD+40~4223 is somewhat higher. 

Inspecting our MCMC posteriors in Figure~\ref{fig:bd404223_corner}, we find that there are strong correlations between the effective temperature of the star ($T_{\rm eff}$), the extinction toward the star ($A_V$), and the overall normalization ($\Omega_*$).
However, these parameters are uncorrelated with the inferred strength of the 1.05\,$\mu$m feature ($\Delta\tau_{1.05}/A_{V}$). 
In addition, despite these degeneracies, the stellar parameters are tightly constrained.

Although the formal significance reported by our best-fit model is $\approx1200\sigma$, this underestimates the level of uncertainty in our constraint.
Indeed, even a model with $\Delta\tau = 0$ is excluded by the data with a significance of $80\sigma$, as evidenced by the bias in the residuals of the best-fit model plotted in Figure~\ref{fig:bd40_105}.
Thus, while the extinguished blackbody model provides a close representation of data, it does not do so at the precision of the high S/N measurements.
To account for the modeling uncertainty in our fit, we quantify the deviation of the best-fit model from the TripleSpec observations at 1.05~$\mu$m, i.e., the level of bias observed in Figure~\ref{fig:bd40_105}.
The residual at 1.05\,$\mu$m is 0.03~Jy, corresponding to a modeling uncertainty of $\Delta\tau_{1.05}/A_{V}=1.1\times10^{-3}$.
Adopting this uncertainty yields an adjusted significance of the non-detection of $\approx 15\sigma$ and a corresponding $5\sigma$ upper limit on the strength of the feature of $\Delta\tau_{1.05}/A_{V} < 5.6 \times10^{-3}$.
As this better reflects the uncertainty in our data and modeling, we adopt $\Delta\tau_{1.05}/A_{V} =(9.5\times10^{-6}) \pm (1.1\times10^{-3})$ for the discussion.

%
%
\section{Discussion}\label{sec:discussion} 
\subsection{
    \texorpdfstring{The Strength of the 1.05~$\mu$m PAH Feature}{The Strength of the 1.05 micron PAH Feature}
}
\label{sec:subsec:no_105_line}

Our non-detection places strict constraints on the strength of the 1.05~$\mu$m~feature.
As our value is $> 10\sigma$ less than the predicted value from (see Figure~\ref{fig:bd40_105}), our work strongly favors the absence of this feature in the spectra of interstellar PAHs.

The possible existence of the 1.05~$\mu$m PAH feature originates in the laboratory work of \citet{2005ApJ...629.1183M, 2005ApJ...629.1188M} where the near-infrared spectra of various PAH cations revealed a weak absorption band centered at approximately 1.05~$\mu$m.
It was then first adopted into PAH dust models in \citet{2007ApJ...657..810D}. 
While some subsequent dust models have continued to include it \citep[e.g.,][]{2021ApJ...917....3D, 2023ApJ...948...55H}, others have opted to omit it~\citep[e.g.,][]{2018A&A...610A..16G}.

To our knowledge, there has been no prior work dedicated to detecting the 1.05~$\mu$m~feature.
However, using the pair method, \citet{2022ApJ...930...15D} investigated the near-infrared extinction curve along multiple sight lines in the $0.8-5.5$~$\mu$m~range using IRTF/SpeX spectra.
While most of the targets were only modestly extinguished by dust, two targets---HD 029647 and HD 283809---were behind 3.52 and 5.65 magnitudes of visual extinction respectively.
Inspecting the observations of these two targets similarly shows no evidence of an absorption feature at 1.05~$\mu$m.

Unlike longer-wavelength PAH features observed at infrared wavelengths (e.g., 3.3, 6.2), the 1.05{~$\mu$m~}feature is not the result of the vibrational modes of PAHs. 
Instead, the 1.05~$\mu$m~feature is believed to originate from the electronic transitions of ionized PAHs, specifically PAH cations~\citep{2005ApJ...629.1183M, 2007ApJ...657..810D}.
One possible reason for the absence of this feature perhaps lies with the population of ionized PAHs.
The absence of the 1.05~$\mu$m~feature could indicate that the majority of PAHs on this line of sight are neutral {as the} electronic transitions of neutral PAHs occur at {shorter wavelengths~\citep{2005ApJ...629.1188M}. 
However,} such a scenario would be at odds with the consensus that ionized PAHs are widespread \cite[e.g.,][]{1996A&A...315L.353M, 1996PASJ...48L..59O}{. Dust models have required a certain fraction of PAHs to be ionized in order to replicate observations \citep[e.g.,][]{2004ASPC..309..665H}.
Should there be more variation in the fraction of ionized versus neutral PAHs than previously assumed, revisions to existing dust models may be required.
}

In a similar vein, as the 1.05~$\mu$m~feature originates from PAH cations specifically, the proportion of cations to anions is also relevant. 
Absorption features due to electronic transitions in PAH anions are also predicted.
Such features would not only be centered at different wavelengths, but they are substantially subdued compared to those from cations~\citep{2005ApJ...629.1183M}.
If the ionized PAH population is largely dominated by anions, this could also explain the lack of the 1.05~$\mu$m~feature.

Alternatively, it is possible that the specific species of ionized PAHs investigated by \citet{2005ApJ...629.1188M} in the laboratory are not representative of the PAHs in the interstellar medium.
In their study where this feature was detected, \citet{2005ApJ...629.1188M} obtained the spectrum for 27 different PAH cations and anions of various sizes (C$_{14}$H$_{10}$ to C$_{50}$H$_{22}$). 
Should these specific molecules not be widely present in the interstellar medium, neither might this feature.

{
Though no prior work has been dedicated to detecting electronic transitions from ionized PAHs~(such as the 1.05~$\mu$m~feature), there have been studies to measure such transitions from neutral PAH molecules. 
In all cases, these efforts have yielded strong upper limits on the presence of electronic transitions. \citet{2003ApJ...592..947C} studied the UV extinction curve using HST/STIS and found no evidence of absorption features from PAHs. Searches for the electronic transitions of specific neutral PAH molecules (e.g., pyrene, acenaphthene, anthracene) using ground based near-UV and optical spectra also resulted in non-detections \citep{2011ApJ...728..154S, 2011A&A...530A..26G}.
Given the lack of electronic transitions from neutral and positively charged PAHs, one implication may be that the PAHs in the interstellar medium predominantly carry negative charge.
However, this is not predicted from physical models of PAHs~\citep[e.g.,][]{2001ApJ...548..296W} that predict small PAHs to be either neutral or cationic and larger PAHs to be predominantly cationic.
On the other hand, it may be that none of the PAH molecules evaluated in laboratory experiments that possess these electronic transitions exist in large quantities in the interstellar medium. 
In both cases, this further suggests the need for additional laboratory experiments to explore the observational signatures of a wider range of charged and neutral hydrocarbons~\citep[for a detailed discussion, see, e.g.,][]{2022Ap&SS.367...16K}.
}

\subsection{Unknown Spectral Features}
\label{sec:subsec:otherspectralfeatures}

\begin{figure}[!t]
    \centering
    \includegraphics[width=1\linewidth]{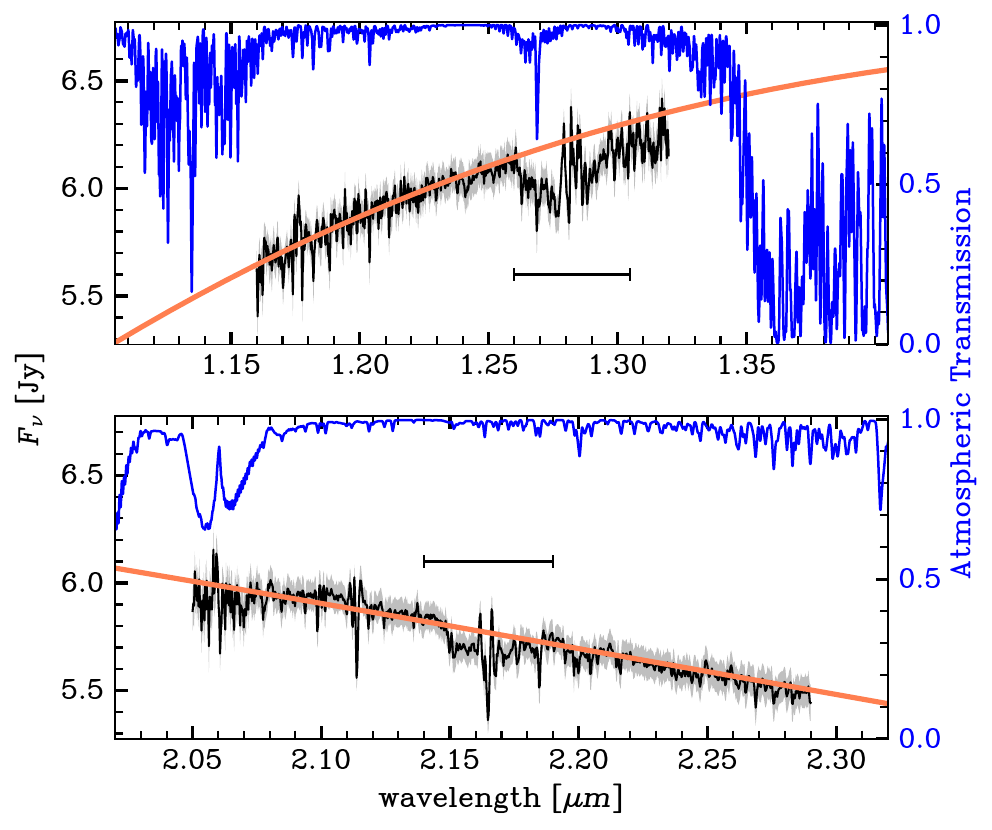}
    \caption{
    Unknown spectral features observed in the TripleSpec data.
    TripleSpec observations are shown in black with $1\sigma$ uncertainties shown in gray. 
    The best-fit model spectrum, fit to the TripleSpec observations, is shown as the orange line. 
    The blue represents the transmission at that wavelength as computed using the ATRAN atmospheric transmission models \citep{1992nstc.rept.....L}.
    The black line segments regions of possible broad absorption features that deviate from our best-fit model spectrum.
    \label{fig:other-features}}
\end{figure}

{While the focus of this work is on the 1.05~$\mu$m feature, other broad features appear in our observations.
Figure~\ref{fig:other-features} shows the TripleSpec observations and the best-fit model near regions where we have identified absorption features that deviate systematically from the best-fit model.
The first is centered at 1.28~$\mu$m~and appears to extend from 1.26 to 1.3~$\mu$m.
This feature coincides with the three strong absorption lines including the Paschen-beta (Pa$\beta$) hydrogen recombination line.
In addition, we also note a broad shallow feature at approximately 2.165~$\mu$m and extending from 2.14 to 2.19~$\mu$m. 
This wavelength range also includes two absorption feature, including the Brackett-gamma (Br$\gamma$) line.
}

{
To our knowledge, neither of these two absorption features have been previously identified as originating from the interstellar medium nor is it clear that these are interstellar in origin.
Indeed, neither feature is seen in the IRTF/SpeX observations from \citet{2022ApJ...930...15D}.
However, both wavelength ranges in these regions do show non-negligible atmospheric absorption~(See Figure~\ref{fig:other-features}).
As a result, we note that it is possible that these features in our observations resulted from an imperfect correction from the atmosphere.
}

\subsection{
    \texorpdfstring{The 1.26~$\mu$m PAH Feature}{The 1.26 Micron PAH Feature}
}\label{subsec:126PAHFeature}
In addition to the 1.05~$\mu$m feature, the \citet{2007ApJ...657..810D} PAH model also includes a feature at 1.26~$\mu$m. 
Like the 1.05~$\mu$m feature, this feature also originates from laboratory measurements of electronic transitions from cationic PAHs.
Compared to the 1.05~$\mu$m feature, the 1.26~$\mu$m feature is expected to be even weaker with \citet{2023ApJ...948...55H} predicting it to be about a quarter of the strength of the 1.05~$\mu$m feature ($\Delta\tau_{1.26}/A_{V} = 0.004$ versus $\Delta\tau_{1.05}/A_{V} = 0.017$).

As the TripleSpec observations includes this wavelength range, in principle, we could have also included the 1.26~$\mu$m feature in our model.
However, as discussed above in Section~\ref{sec:subsec:otherspectralfeatures}, the wavelength range around 1.26~$\mu$m is dominated by a strong unidentified feature in absorption centered at 1.28~$\mu$m.
Given that the 1.26~$\mu$m~feature is predicted to be much weaker by comparison (i.e., $\Delta\tau_{1.26}/A_{V} = 0.004$ versus $\Delta\tau_{1.05}/A_{V} = 0.017$), we would not be able to obtain constraints on its presence.

\subsection{The 770 and 850~nm Features}
\label{subsec:Gaia770850Feature}
The Gaia XP spectrum obtained toward BD+40~4223 also show two extinction features centered at approximately 770~nm and 850~nm~\citep{2021MNRAS.501.2487M, 2025ApJ...988....5G, 2025arXiv250707162S}.
These features are ubiquitous in the Galactic interstellar medium and, while their exact origin is not known, there is evidence linking them to small carbonaceous nanoparticles~\citep{2021MNRAS.501.2487M, 2025ApJ...988....5G}{.}
Computing the equivalent widths~(EW) of the 770~nm and 850~nm feature toward BD+40~4223, we find $EW_{770}/E(B-V) = 2.5$~nm and $EW_{850}/E(B-V) = 1.8$~nm.
These are consistent with the typical values measured using Gaia observations for $\sim 28$ million stars~\citep{2025Sci...387.1209Z, 2025ApJ...988....5G}.
For comparison, the EW per $E$ extinction measurements for both features from \citet{2025ApJ...988....5G} peak at approximately $\sim1$~nm with a long tail of measurements out to $4.5$~nm.

\subsection{Caveats and Interpretation}
Though our analysis strongly implies a lack of a 1.05~$\mu$m PAH feature, we examine several potential concerns. 

\subsubsection{Telluric Correction and Flux Calibration}
We consider whether an improper telluric correction and/or flux calibration may inhibit detection of the 1.05\,$\mu$m feature.
As discussed in Section~\ref{sec:observations}, the telluric correction and flux calibration was completed simultaneously using the Spextool software. 
This procedure uses a telluric standard, HD~201320 in this case, to correct for atmospheric contamination.

Overall, the wavelength range at 1.05\,$\mu$m is a region of exceptional atmospheric transmission.
Using the ATRAN atmospheric transmission models \citep{1992nstc.rept.....L}, and examining the wavelength range between 1.00\,$\mu$m~and 1.07\,$\mu$m, the transmission is near complete ($>0.999$) with no visible telluric features or lines. 
In general, no atmospheric feature coincides with the expected width and depth of the 1.05\,$\mu$m feature.

Similarly, we find no evidence for issues with the flux calibration. 
As shown in Figure~\ref{fig:all-data}, the reduced TripleSpec data are in good agreement with observations from Gaia and 2MASS.
As such we consider it unlikely that the telluric correction or flux calibration erroneously inhibited detection of the PAH feature. 

\subsubsection{Stellar Atmosphere Fits}
We next consider whether an extinguished blackbody is a poor approximation for the type of massive stars targeted in this study.
Overall, we find that our extinguished blackbody model is able to fit the near-infrared spectrum well.
Qualitatively, Figure~\ref{fig:all-data} shows that our model is in good agreement with the measured TripleSpec measurements across the wavelength range.
In addition, we find agreement between our fit and the Gaia XP optical spectrum in Figure~\ref{fig:all-data}---despite not being fit to it---suggesting no obvious problem with using an extinguished blackbody to fit the observations. 

In principle, we could have used theoretical stellar atmosphere models in place of a blackbody.
However, we do not find this to be necessary.
Examining the TLUSTY OSTAR2002 and BSTAR2006 grid consisting of O-type and B-type star stellar atmospheres~\citep{1995ApJ...439..875H, 2003ApJS..146..417L, 2007ApJS..169...83L}, we explored the impact of fitting a simple blackbody to these more complicated model atmospheres.
Including a range of values in effective temperature, surface gravity, and metallicity, we found that the fits to the models underestimated the true effective temperature and extinction yet still provided a good fit. Indeed, this could potentially explain our higher than expected estimate of BD+40~4223's radius described in Section~\ref{sec:results}. 
The effective temperature and extinction underestimates would be absorbed by our normalization constant. 
Assuming a well-measured distance from Gaia parallax, this would inflate our computed stellar radius.

While more complex stellar atmosphere models could result in more accurate estimates of the stellar parameters or the overall level of extinction, neither of these are factors are limiting constraints on the strength of the 1.05~$\mu$m feature, and indeed have essentially no degeneracy with it (see Figure~\ref{fig:bd404223_corner}).
Furthermore, it has been noted that these theoretical models suffer from a combination of unreliable continuum level and an incomplete inventory of stellar lines at longer wavelengths~\citep[][]{2020ApJ...891...67M, 2022ApJ...930...15D}.
As such, incorporating these theoretical models may not necessarily be a more accurate approach.

\subsubsection{Local Dust Properties}
Since our observations only target a single sight line toward Cyg~OB2, it is possible that something atypical along this sight line prevents detection of the PAH feature at 1.05\,$\mu$m.
Specifically, if our sight line is uniquely not representative of the typical Galactic interstellar medium in some manner (such as for the reasons described in Section~\ref{sec:subsec:no_105_line}), we would, by chance, unfortunately not detect the 1.05~$\mu$m~feature.
While only observations toward other sight lines will truly be able to determine whether this is true or not, there are several lines of evidence that suggest that the sight line toward BD+40~4223 is typical of the Galactic interstellar medium.

First, observations toward Cyg OB2-12, a nearby star also in the Cyg~OB2 association, show a typical extinction curve and lack of strong ice features~\citep{1997ApJ...490..729W, 2015ApJ...811..110W}.
This suggests that the bulk of the extinction originates from dust in the diffuse interstellar medium rather than dense gas that can have a more peculiar composition.
Given that BD+40 4223 is located in the same association, it would be unusual for the two sight lines in the same region to differ significantly.
As a result, we expect the sight line toward BD+40~4223 to also to be dominated by dust typical of the diffuse interstellar medium.

It is possible that the lack of the 1.05~$\mu$m~feature is due to the lack of PAHs in general toward Cyg~OB2.
However, analysis of archival Infrared Space Observatory and Spitzer observations---again toward Cyg OB2-12---prominently show the 3.3, 6.2, and 7.7~$\mu$m PAH features in absorption~\citep{2020ApJ...895...38H}.
Given that these features were readily detected, it is unlikely that a general lack of PAHs toward Cyg~OB2 can explain the non-detection of the 1.05~$\mu$m~feature.

Finally, the Gaia XP spectra toward BD+40~4223 show the 770~nm and 850~nm feature with equivalent widths of 2.5~nm and 1.8~nm, respectively. 
These measurements fall squarely into the range of values measured by \citet{2025ApJ...988....5G} ($\sim1$ to 4.5 nm) for $24$ million stars.
The presence and strength of the 770~nm and 850~nm features is thus further evidence that the dust composition toward BD+40~4223 similar to other sight lines across the sky.

As a result, while only with multiple observations will we truly know for certain, given the lines of evidence that the dust toward BD+40~4223 is fairly typical, we argue that our strong non-detection of the 1.05~$\mu$m feature is representative of the  Galactic diffuse interstellar medium.

\section{Conclusion}\label{sec:conclusion}

In this work, we target the heavily extinguished sight line toward the massive star BD+40~4223 with the TripleSpec spectrograph to search for evidence of the 1.05~$\mu$m feature in absorption.
This feature is believed to originate from the weak electronic transitions of PAHs.
Modeling our observations we derive strict constraints on the presence of the 1.05~$\mu$m PAH feature.

Our primary results are as follows:
\begin{enumerate}
    \item We report a non-detection of the 1.05~$\mu$m feature with a $5\sigma$ upper limit of $\Delta\tau_{1.05}/A_{V} < 5.6 \times10^{-3}$. This is $> 10\sigma$~inconsistent with theoretical estimates of the feature strength~\citep[$\Delta\tau_{1.05}/A_{V}=0.017$,][]{2023ApJ...948...55H}.
    Our non-detection toward BD+40~4223 is consistent with IRTF/SpeX spectra of sight lines toward HD~029647 and HD~283809~\citep{2022ApJ...930...15D}, which both show no evidence of a 1.05~$\mu$m absorption feature. 

    \item Our TripleSpec observations show unknown spectral features at $\sim1.28$~$\mu$m and $\sim2.165$~$\mu$m. Both coincide with strong hydrogen recombination lines, are in regions with non-negligible atmospheric absorption, and are not seen in the IRTF/SpeX spectra of HD~029647 and HD~283809, and so are unlikely to be true interstellar dust features.

    \item The 1.26~$\mu$m PAH feature---also predicted to originate from electronic transitions of PAH cations---lies in the wavelength range observed with TripleSpec. However, the unknown $\sim1.28$~$\mu$m spectral feature as well as the appreciable atmospheric absorption inhibits constraints on its strength using our data.

    \item We detect the 770 and 850~nm features in the Gaia XP spectra of BD+40~4223. 
    The strengths of both features are consistent with the strength of the 770 and 850~nm feature measured for 24 million stars across the sky~\citep{2025ApJ...988....5G}.

    \item Since the 1.05~$\mu$m feature is from cationic PAHs, {its} non-detection poses challenges for our current models of interstellar dust. {Taking our results here together with non-detections of neutral electronic transitions as well as other deviations from the typical PAH models \citep[e.g.,][]{2005ApJ...632..316H, 2015ApJ...807...95S, 2021ApJ...916...52T,2025MNRAS.544.3280T}, this} suggests revisions may be required to our assumptions on the charge distribution of PAH molecules (i.e., ionized versus neutral, cationic versus anionic) or {the chemical nature of the molecules themselves}. 

    \item Our non-detection of the 1.05~$\mu$m feature does not preclude the existence of other spectral features from PAH electronic transitions.
    Since the 1.05~$\mu$m feature is believed to originate from PAH cations, electronic transitions from neutral PAHs or PAH anions may exist at other near-infrared wavelengths~{(though, despite dedicated searches, neutral electronic transitions have yet to be confirmed)}.
    Additionally, the cationic species that have been measured in the laboratory may not be representative of the PAH cations present in the interstellar medium, which could have electronic transitions at other wavelengths.
    We nevertheless find no compelling evidence for other interstellar features on the sight line toward BD+40~4223.

\end{enumerate}

Although the sight line toward BD+40~4223 seems prototypical of the diffuse interstellar medium, we emphasize that follow-up work is yet required to diagnose whether or not the feature exists in astrophysical environments.
The 1.05~$\mu$m feature falls in the bands of JWST, SPHEREx, as well as various ground-based near-infrared spectrographs.
Observations with such instruments and facilities toward other parts of our Galaxy---as well as those in other galaxies---are vital to assessing whether the 1.05~$\mu$m feature, or other features arising from electronic transitions, are present in the spectra of interstellar PAHs. 

\begin{acknowledgments}
{
We thank the anonymous referee for providing helpful comments that improved the manuscript.
We thank S. Rose for the advice on the reduction of the TripleSpec data.
We thank A. Saydjari, X. Zhang, and G. Green for their assistance with accessing the Gaia XP data.
We thank E. Mamajek and the JPL/IPAC Palomar TAC, the staff at Palomar Observatory, and F.~Marocco for their assistance in obtaining the TripleSpec observations.}
This research was carried out at the Jet Propulsion Laboratory, California Institute of Technology, under a contract with the National Aeronautics and Space Administration (80NM0018D0004). 
Based on observations obtained at the Hale Telescope, Palomar Observatory, as part of a collaborative
agreement between the Caltech Optical Observatories and the Jet Propulsion Laboratory [operated by Caltech
for NASA].
\end{acknowledgments}
\facility{Hale (TripleSpec)}

\software{
    \texttt{astropy} \citep{2013A&A...558A..33A, 2018AJ....156..123A, 2022ApJ...935..167A},
    \texttt{numpy} \citep{harris2020array}, 
    \texttt{scipy} \citep{2020NatMe..17..261V}, 
    \texttt{matplotlib} \citep{2007CSE.....9...90H},
    \texttt{emcee} \citep{2013PASP..125..306F},
    \texttt{corner.py} \citep{2016JOSS....1...24F}
    \texttt{dust\_extinction} \citep{2024JOSS....9.7023G}
}

\bibliography{reference, Planck_bib, software}{}
\bibliographystyle{aasjournalv7}
\end{document}